\documentclass[10pt,a4paper]{article}

\title{Construction of Gray maps for groups of order 16}

\author{Ko Sakai$^{1}$ and Yutaka Sato$^{2}$\\
$^{1}$Faculty of Pure and Applied Sciences,\\
$^{2}$Graduate School of Pure and Applied Sciences,\\
 University of Tsukuba, Tsukuba city, Japan}

\usepackage{amsmath,amssymb}

\newtheorem{thm}{Theorem}
\newtheorem{lmm}{Lemma}
\newtheorem{dfn}{Definition}

\begin{document}

\maketitle

Abstract. In this paper we consider  Type 1 Gray maps and  Type 2 Gray maps for groups of order 16.  First, we confirm that we can construct  Type 1 Gray maps for all groups of order 16, and we actually construct  them. Next, we confirm that we can construct Type 2 Gray maps for several groups of order 16, and we construct all such maps. Finally we reveal why the other groups do not admit Gray maps. 
 
\section{Introduction}
 Reza Sobhani constructed two classes of Gray maps, called Type 1 Gray map and Type 2 Gray map, for finite $p$-groups $G$ in [1].
Type 1 Gray maps are constructed based on a Gray map for a maximal subgroup $H$ of $G$. 
In contrast, Type 2 Gray maps are constructed, if $G$ is a semidirect product of two finite $p$-group $H$ and $K$, both $H$ and $K$ admit Gray maps and the corresponding homomorphism $\psi : H \rightarrow$ Aut($K$) is compatible with the Gray map for $K$ in the sense that for all $h \in H$ we have $w_{H}(\theta_{2}(k))$ = $w_{H}(\theta_{2}(\psi_{h}(k)))$. 
Type 2 Gray maps have feature which they generally generate shorter codes than Type 1 Gray maps. However, we can construct for only 6 groups among all the groups of order 16.   

 Marcel Wild gave complete classification of the groups of order 16 based on several elementary facts. He examined them as  extensions of groups of order 8 by the cyclic group of order 2. 

 In this paper we consider Type 1 Gray maps and Type 2 Gray maps for groups of order 16 in the same line as Sobhani. We assume that information is encoded in an alphabet $Q$ with $q$ distinct symbols.

\section{Preliminaries}
In this section we assume that $G$ is a finite 2-group of order $2^m$. 
\subsection{Hamming-distance, Hamming-weight and Gray map}

\begin{dfn}\rm 
For any two elements $\textbf{u} = (u_{1}, u_{2}, \dots ,u_{n})$ and $\textbf{v} = (v_{1}, v_{2}, \dots ,v_{n})$ in $Q^n$, the \textit{Hamming-distance} between $\textbf{u}$ and $\textbf{v}$ is defined by,

$d_{H}(\textbf{u},\textbf{v}) \stackrel{\mathrm{def.}}{=} |\{i \mid 1 \le i \le n, u_{i} \neq v_{i} \} |$.
\end{dfn}
\begin{dfn}\rm
The \textit{Hamming-weight} of $\textbf{u}$ is defined by,

$w_{H}(\textbf{u}) \stackrel{\mathrm{def.}}{=} |\{i \mid 1\le i \le n, u_{i} \neq 0 \}|$.
\end{dfn}
We review the definition and a lemma on a Gray map in [1].

\begin{dfn}\rm
A map $\phi : G \rightarrow \mathbb{Z}^n_{2}$ is said to be a \textit{Gray map}, if the following properties hold, \\
(1) The map $d_{\phi} : G \times G \rightarrow \mathbb{N} \cup \{0\}$ defined by  $d_{\phi}(a,b) = w_{H}(\phi(ab^{-1}))$ is a distance on $G$. \\
(2) For all $a,b$ in $G$ we have $d_{\phi}(a,b) = d_{H}(\phi(a),\phi(b))$.
\end{dfn}

\begin{lmm}\rm
 Condition (1) in the definition of a Gray map is equivalent to the following conditions, \\
(1) For $g \in G$ we have $w_{H}(\phi(g))$ = 0 iff $g = e$, where $e$ stands for the identity of $G$, \\
(2) For all $g$ in $G$ we have $w_{H}(\phi(g))$ = $w_{H}(\phi(g^{-1}))$, \\
(3) For all $x,y$ in $G$ we have $w_{H}(\phi(xy)) \le w_{H}(\phi(x)) + w_{H}(\phi(y))$. 
\end{lmm}
Refer to in [1] for the proof of Lemma 1.

$\textbf{Example 2.1.}$ Gray map over $C_{4} = \langle x \mid x^4 = e \rangle$.  

$\textbf{Example 2.2.}$ Gray map over $K_{4} = \langle x,y \mid x^2 = y^2 = e, xy = yx \rangle$.  (We write $K_{4}$ for the \textit{Klein four group}.) Gray maps as follows.

\begin{center}
 \begin{tabular}{ll|ll}
   \multicolumn{4}{c}{\textbf{Example 2.1 and 2.2}} \\ \hline\hline
    $C_{4}$ & $\mathbb{Z}^2_{2}$ & $K_{4}$ & $\mathbb{Z}^2_{2}$ \\ \hline
    $\phi(e)$ & 00      & $\phi(e)$ & 00 \\
    $\phi(x)$ & 01      & $\phi(x)$ & 01 \\ 
    $\phi(x^2$) & 11   & $\phi(y)$ & 11 \\   
    $\phi(x^3$) & 10   & $\phi(xy)$ & 10 \\ \hline
 \end{tabular}
\end{center}

\subsection{Type 1 Gray maps}
In this subsection, we assume that $H$ is a  maximal subgroup of $G$ with [$G : H$] = 2, and $x$ is an arbitrary elements in $G \setminus H$ and $h$ is an arbitrary elements in $H$.  In this part, we construct a Gray map for $G$, called Type 1 Gray map, based on the existence of a Gray map for $H$. For 1 $\in \mathbb{Z}_{2}$, let us denote by \textbf{1} the vector in  $\mathbb{Z}^n_{2}$ whose components are all 1. Also we denote the usual concatenation of vectors in $\mathbb{Z}^n_{2}$ by $( \              \mid \ )$. Suppose $\phi : H \rightarrow \mathbb{Z}^n_{2}$ is a Gray map and define the map $\hat{\phi} : G \rightarrow \mathbb{Z}^{2n}_{2}$ by
 $\hat{\phi}(h) = (\phi(h) \mid \phi(h))$ and $\hat{\phi}(xh) = (\phi(h) \mid \phi(h) + \textbf{1})$ in [1].

\begin{lmm}\rm
 For all $g \in G$ we have $w_{H}(\hat{\phi} (g)) = w_{H}(\hat{\phi} (g^{-1}))$. 
\end{lmm}

\begin{lmm}\rm
 For all $a,b \in G$ we have $w_{H}(\hat{\phi} (ab)) \le w_{H}(\hat{\phi}(a)) + w_{H}(\hat{\phi}(b))$. 
\end{lmm}

\begin{thm}\rm
 With notation as above, the map $\hat{\phi}$ is a Gray map. \end{thm}

Refer to in [1] for the proof of Lemma 2, 3 and Theorem 1.

\subsection{Type 2 Gray maps}
In this subsection, we assume that $G$ is isomorphic to the semidirect product of two finite 2-groups $H$ and $K$ of orders  $2^a$ and $2^b$ respectively, i.e. $G = H \ltimes_{\psi} K$ where $\psi : H \rightarrow$ Aut($K$) is a group homomorphism. Let $\theta_{1} : H \rightarrow \mathbb{Z}^{n_{1}}_{2}$ and $\theta_{2} : K \rightarrow \mathbb{Z}^{n_{2}}_{2}$ be Gray maps, where $\theta_{2}$ is compatible with $\psi$ in the sense that for all $h \in H$ 
\[ w_{H}(\theta_{2}(k)) = w_{H}(\theta_{2}(\psi_{h}(k))). \]
Then we show that $\theta : G \rightarrow \mathbb{Z}^{n_{1} + n_{2}}_{2}$ with $\theta(hk) = (\theta_{1}(h) \mid \theta_{2}(k))$ is a Gray map,  which we call Type 2 Gray map.

\begin{thm}\rm
 With notation as above, if  $\theta_{2}$ is compatible with $\psi$, then the map $\theta$ is a Gray map. 
\end{thm}

Refer to in [1] for the proof of Theorem 2.

\subsection{Automorphism}
The set Aut($G$) of all automorphisms of a group $G$ forms a group under composition of mappings. Let $X$ generate $G$. Each  $\theta : G \rightarrow G$ in Aut($G$) is determined by its values on $X$.
For notational convenience we view the direct product of  $C_{4}  = \langle x \rangle$ and  $C_{2} = \langle y \rangle$ as \{e, $x, x^2 , x^3, y, xy, x^2y, x^3y$\}. Because this group will appear frequently in this article we denote it by $K_{8}$ [2]. Similarly, 
 $D_{8} = \langle x,y \mid x^4 = y^2 = e, yx = x^3y \rangle$ is the dihedral group of order 8 and $Q_{8} = \langle x,y \mid x^4 = e,y^2 = x^2, yx = x^3y \rangle$ is the quaternion group of order 8.

$\textbf{Example 2.3.}$ 
\begin{center}
 \begin{tabular}{ll|ll}
   \multicolumn{4}{c}{\textbf{Aut($C_{4}$) and Aut($C_{8}$)$\simeq$$K_{4}$} [2]} \\ \hline\hline
   Aut($C_{4}$) & effect on $x$ & Aut($C_{8}$) & effect on $x$ \\ \hline
    $\varphi_{1}$  & $x$    & $\sigma_{1}$  & $x$ \\
    $\varphi_{2}$  & $x^3$ & $\sigma_{2}$  & $x^3$ \\ 
                  &         & $\sigma_{3}$  & $x^5$ \\   
                  &         & $\sigma_{4}$  & $x^7$ \\ \hline
 \end{tabular}
\end{center}
 
$\textbf{Example 2.4.}$ 
\begin{center}
 \begin{tabular}{llll}
   \multicolumn{4}{c}{\textbf{Aut($K_{8})\simeq D_{8}$} [2]} \\  \hline\hline
    Aut($K_{8}$) &effect on $x$ & effect on $y$ & order of automorphism \\ \hline
    $\psi_{1}$ & $x$        & $y$  & 1 \\
    $\psi_{2}$ & $x^3y$   & $x^2y$ & 4 \\ 
    $\psi_{3}$ & $x^3$     & $y$ & 2 \\   
    $\psi_{4}$ & $xy$      & $x^2y$ & 4 \\ 
    $\psi_{5}$ & $xy$      & $y$ & 2 \\ 
    $\psi_{6}$ & $x^3$     & $x^2y$ & 2 \\ 
    $\psi_{7}$ & $x^3y$   & $y$ & 2 \\ 
    $\psi_{8}$ & $x$       & $x^2y$ & 2 \\  \hline
 \end{tabular}
\end{center}

$\textbf{Example 2.5.}$ 
\begin{center}
 \begin{tabular}{llll}
   \multicolumn{4}{c}{\textbf{Aut($D_{8}$)$\simeq$$D_{8}$}} \\  \hline\hline
    Aut($D_{8}$) &effect on $x$ & effect on $y$ & order of automorphism \\ \hline
    $\alpha_{1}$ & $x$    & $y$  & 1 \\
    $\alpha_{2}$ & $x$    & $xy$ & 4 \\ 
    $\alpha_{3}$ & $x$    & $x^2y$ & 2 \\   
    $\alpha_{4}$ & $x$    & $x^3y$ & 4 \\ 
    $\alpha_{5}$ & $x^3$ & $y$ & 2 \\ 
    $\alpha_{6}$ & $x^3$ & $xy$ & 2 \\ 
    $\alpha_{7}$ & $x^3$ & $x^2y$ & 2 \\ 
    $\alpha_{8}$ & $x^3$ & $x^3y$ & 2 \\  \hline
 \end{tabular}
\end{center}

\section{Type 1 Gray maps for a group of order 8}
Let $G$ be $C_{4} = \langle x \rangle$. Assume that $H = \{e, x^2\} \le G$ be the maximal subgroup of $G$, which is isomorphic to $C_{2}$. Let $\phi_{0} : H \rightarrow \mathbb{Z}_{2}$ be the natural map which sends $e$ to 0 and $x^2$ to 1. Clearly $\phi_{0}$ is a Gray map. Set $\phi_{1} \stackrel{\mathrm{def.}}{=} \hat{\phi}_{0}$, we have  $\phi_{1}(e) = (\phi_{0}(e) \mid \phi_{0}$(e)) = 00, $\phi_{1}(x^2) = (\phi_{0}(x^2) \mid \phi_{0}(x^2))$ = 11, $\phi_{1}(x) = (\phi_{0}(e) \mid (\phi_{0}$(e) + 1)) = 01 and $\phi_{1}(x^3) = (\phi_{0}(x^2) \mid (\phi_{0}(x^2)$ + 1)) = 10. This is the well-known Gray map on $C_{4}$ in [1].

$\textbf{Example 3.1.}$ Let $G$ be $C_{8} = \langle x \rangle$. Assume that $H = \{e, x^2, x^4, x^6\} \le G$ be the maximal subgroup of $G$, which is isomorphic to $C_{4}$. Let $\phi_{1} : H \rightarrow \mathbb{Z}^2_{2}$ be the previously constructed Gray map for $C_{4}$. We have,
\begin{center}
 \begin{tabular}{lll}
  $\hat{\phi}_{1}(e)$ & = ($\phi_{1}(e)\mid \phi_{1}(e)$) &= 0000 \\
 $\hat{\phi}_{1}(x^2)$ & = ($\phi_{1}(x)\mid \phi_{1}(x)$) &= 0101 \\ 
$\hat{\phi}_{1}(x^4)$&= ($\phi_{1}(x^2)\mid \phi_{1}(x^2)$)&= 1111\\   
$\hat{\phi}_{1}(x^6)$&= ($\phi_{1}(x^3)\mid \phi_{1}(x^3)$)&= 1010\\ 
$\hat{\phi}_{1}(x)$&= ($\phi_{1}(e)\mid \phi_{1}(e)$ + 11)&= 0011\\ 
$\hat{\phi}_{1}(x^3)$&= ($\phi_{1}(x)\mid \phi_{1}(x)$ + 11)&= 0110  \\ 
$\hat{\phi}_{1}(x^5)$& = ($\phi_{1}(x^2)\mid \phi_{1}(x^2)$ + 11) &= 1100 \\ 
$\hat{\phi}_{1}(x^7)$& = ($\phi_{1}(x^3)\mid \phi_{1}(x^3)$ + 11) &= 1001 \\  
 \end{tabular}
\end{center}  

$\textbf{Example 3.2.}$ Let $G$ be $K_{8}$. Similarly, assume that $H = \{e, x, x^2, x^3\} \le G$ be the maximal subgroup of $G$. Let $\phi_{1} : H \rightarrow \mathbb{Z}^2_{2}$ be the previously constructed Gray map for $C_{4}$. We have,
\begin{center}
 \begin{tabular}{lll}
 $\hat{\phi}_{1}(e)$ & = ($\phi_{1}(e)\mid \phi_{1}(e)$) &= 0000 \\
 $\hat{\phi}_{1}(x)$ & = ($\phi_{1}(x)\mid \phi_{1}(x)$) &= 0101 \\ 
$\hat{\phi}_{1}(x^2)$&= ($\phi_{1}(x^2)\mid \phi_{1}(x^2)$)&= 1111\\   
$\hat{\phi}_{1}(x^3)$&= ($\phi_{1}(x^3)\mid \phi_{1}(x^3)$)&= 1010\\ 
$\hat{\phi}_{1}(y)$& = ($\phi_{1}(e)\mid \phi_{1}(e)$ + 11)&= 0011\\ 
$\hat{\phi}_{1}(xy)$&= ($\phi_{1}(x)\mid \phi_{1}(x)$ + 11)&= 0110\\ 
$\hat{\phi}_{1}(x^2y)$&= ($\phi_{1}(x^2)\mid \phi_{1}(x^2)$ + 11) &= 1100 \\ 
$\hat{\phi}_{1}(x^3y)$ & = ($\phi_{1}(x^3)\mid \phi_{1}(x^3)$ + 11) &= 1001 \\  
 \end{tabular}
\end{center}  

$\textbf{Example 3.3.}$ Let $G$ be $D_{8}$. Similarly, assume that $H = \{e, x, x^2, x^3\} \le G$ be the maximal subgroup of $G$.  Let $\phi_{1} : H \rightarrow \mathbb{Z}^2_{2}$ be the previously constructed Gray map for $C_{4}$. We have,
\begin{center}
 \begin{tabular}{lll}
 $\hat{\phi}_{1}(e)$  & = ($\phi_{1}(e)\mid \phi_{1}(e)$) &= 0000 \\
 $\hat{\phi}_{1}(x)$  & = ($\phi_{1}(x)\mid \phi_{1}(x)$) &= 0101 \\ 
 $\hat{\phi}_{1}(x^2)$ & = ($\phi_{1}(x^2)\mid \phi_{1}(x^2)$) &= 1111 \\   
 $\hat{\phi}_{1}(x^3)$ & = ($\phi_{1}(x^3)\mid \phi_{1}(x^3)$) &= 1010 \\ 
 $\hat{\phi}_{1}(y)$ & = ($\phi_{1}(e)\mid \phi_{1}(e)$ + 11) &= 0011 \\ 
 $\hat{\phi}_{1}(yx)$ & = ($\phi_{1}(x)\mid \phi_{1}(x)$ + 11) &= 0110  \\ 
 $\hat{\phi}_{1}(yx^2)$ & = ($\phi_{1}(x^2)\mid \phi_{1}(x^2)$ + 11) &= 1100 \\ 
 $\hat{\phi}_{1}(yx^3)$ & = ($\phi_{1}(x^3)\mid \phi_{1}(x^3)$ + 11) &= 1001 \\  
 \end{tabular}
\end{center}  

$\textbf{Example 3.4.}$ Let $G$ be $Q_{8}$. Similarly, assume that $H = \{e, x, x^2, x^3\} \le G$ be the maximal subgroup of $G$.  Let $\phi_{1} : H \rightarrow \mathbb{Z}^2_{2}$ be the previously constructed Gray map for $C_{4}$. We have,
\begin{center}
 \begin{tabular}{lll}
 $\hat{\phi}_{1}(e)$    & = ($\phi_{1}(e)\mid \phi_{1}(e)$) &= 0000 \\
 $\hat{\phi}_{1}(x)$    & = ($\phi_{1}(x)\mid \phi_{1}(x)$) &= 0101 \\ 
 $\hat{\phi}_{1}(x^2)$ & = ($\phi_{1}(x^2)\mid \phi_{1}(x^2)$) &= 1111 \\   
 $\hat{\phi}_{1}(x^3)$ & = ($\phi_{1}(x^3)\mid \phi_{1}(x^3)$) &= 1010 \\ 
 $\hat{\phi}_{1}(y)$    & = ($\phi_{1}(e)\mid \phi_{1}(e)$ + 11) &= 0011 \\ 
 $\hat{\phi}_{1}(yx)$   & = ($\phi_{1}(x)\mid \phi_{1}(x)$ + 11) &= 0110  \\ 
 $\hat{\phi}_{1}(yx^2)$ & = ($\phi_{1}(x^2)\mid \phi_{1}(x^2)$ + 11) &= 1100 \\ 
 $\hat{\phi}_{1}(yx^3)$ & = ($\phi_{1}(x^3)\mid \phi_{1}(x^3)$ + 11) &= 1001 \\  
 \end{tabular}
\end{center}  

\section{Type 2 Gray maps for a group of order 8}
$\textbf{Example 4.1.}$ 
Let $G$ be $D_{8}$. We have $G\simeq C_{2} \ltimes_{\psi} C_{4}$, where $\psi : C_{2} \rightarrow$ Aut($C_{4}$) is a homomorphism. Consider the natural Gray map $\phi_{0} : C_{2} \rightarrow \mathbb{Z}_{2}$ and the Gray map $\phi_{1} : C_{4} \rightarrow \mathbb{Z}^2_{2}$ described in the example 2.1. Now we can construct a Type 2 Gray map $\theta$ from $D_{8}$ to $\mathbb{Z}^3_{2}$ as follows in [1],
\begin{center}
 \begin{tabular}{lll}
 $\theta(e)$    & = ($\phi_{0}(e)\mid \phi_{1}(e)$) &= 000 \\
 $\theta(x)$    & = ($\phi_{0}(e)\mid \phi_{1}(x)$) &= 001 \\ 
 $\theta(x^2)$ & = ($\phi_{0}(e)\mid \phi_{1}(x^2)$) &= 011 \\   
 $\theta(x^3)$ & = ($\phi_{0}(e)\mid \phi_{1}(x^3)$) &= 010 \\ 
 $\theta(y)$    & = ($\phi_{0}(y)\mid \phi_{1}(e)$) &= 100 \\ 
 $\theta(yx)$   & = ($\phi_{0}(y)\mid \phi_{1}(x)$) &= 101 \\ 
 $\theta(yx^2)$ & = ($\phi_{0}(y)\mid \phi_{1}(x^2)$) &= 111 \\ 
 $\theta(yx^3)$ & = ($\phi_{0}(y)\mid \phi_{1}(x^3)$) &= 110 \\     
 \end{tabular}
\end{center}
We denote by $\phi_{2}$ the above Gray map $\theta$.

$\textbf{Example 4.2.}$ Similarly we can construct a Type 2 Gray map $\theta$ from $K_{8}$ to $\mathbb{Z}^3_{2}$,
\begin{center}
 \begin{tabular}{llll}
    $\theta(e)$ & = 000       & $\theta(y)$  & = 100 \\
    $\theta(x)$ & = 001       & $\theta(xy)$ & = 101 \\ 
    $\theta(x^2$) & = 011 & $\theta(x^2y)$ & = 111 \\   
    $\theta(x^3$) & = 010 & $\theta(x^3y)$ & = 110 \\  
 \end{tabular}
\end{center}
Also, we denote by $\phi_{2}$ the above Gray map $\theta$.

Since we have neither $C_{8}\simeq C_{2} \ltimes_{\psi} C_{4}$ nor $C_{8}\simeq C_{2} \ltimes_{\psi} K_{4}$,  we can not construct a Type 2 Gray map for $C_{8}$.  

 $\textbf{Example 4.3.}$ Let G be $Q_{8}$.
\begin{center}
 \begin{tabular}{llll}  
    $\theta(e)$ & = 000       & $\theta(y)$  & = 100 \\
    $\theta(x)$ & = 001       & $\theta(yx)$ & = 101 \\ 
    $\theta(x^2$) & = 011 & $\theta(yx^2)$ & = 111 \\   
    $\theta(x^3$) & = 010 & $\theta(yx^3)$ & = 110 \\  
 \end{tabular}
\end{center}
In $Q_{8}$, we have $x^{-1} = x^{3}, y^{-1} = yx^2, (yx)^{-1} = yx^{3}$, so we have $w_{H}(\phi(g)) \neq w_{H}(\phi(g^{-1}$)). Since $\theta$ does not satisfy (2) of Lemma 1, we can not construct a Type 2 Gray map for $Q_{8}$.

In $\mathbb{Z}^3_{2}$ weight distribution is as follows:
\begin{center}
 \begin{tabular}{l|llll}   
   Hamming-weight & 0 & 1 & 2 & 3 \\ \hline
   number of points & 1 & 3 & 3 & 1 \\ 
 \end{tabular}
\end{center}
Since both 1 and 3 are odd, in order to satisfy (2) of Lemma 1, the group $G$ must have at least 3 elements of order 2.

On the other hand, $C_{8}$ has one element of order 2, therefore  can not satisfy (2) of Lemma 1. Similarly, neither does $Q_{8}$. 

\section{Type 1 Gray map for a group of order 16}
\subsection{Cyclic extensions.}
We state the classification of groups of order 16.
Let $N \vartriangleleft G$ and $N^{\prime} \vartriangleleft G^{\prime}$. Suppose that $G/N \simeq C_{n}$.(Such a group $G$ is called an \textit{(inner) cyclic extension} of $N$.) Pick any $a$ in $G \setminus N$ such that the coset $Na$ has order $n$ in $G/N$. Then $v = a^{n}$ is in $N$, and $n$ is minimal with that property. Further, let $\tau$ in Aut($N$) be the restriction of $t_{a}(x) \stackrel{\mathrm{def.}}{=} axa^{-1}$ to $N$. Then 

$\tau(v) = aa^{n}a^{-1} = a^{n} = v$, and

 $\tau^{n} = a\cdot \cdot \cdot a(axa^{-1})a^{-1}\cdot \cdot \cdot  a^{-1} = a^{n}xa^{n} = vxv^{-1} = t_{v}(x)$, for all $x$ in N,

so $\tau^{n} = t_{v}$.
 
\begin{dfn}\rm
 A quadruple ($N, n, \tau, v$) is an \textit{ extension type} if $N$ is a group and if $v$ in $N$ and $\tau$ in Aut($N$) are such that $\tau(v) = v$ and $\tau^n = t_{v}$. 
\end{dfn}

\begin{dfn}\rm
 The extension types ($N, n, \tau, v$) and ($N^{\prime}, n, \sigma, w$) are \textit{equivalent} if there is an isomorphism 
$\phi : N \rightarrow N^{\prime}$ such that $\sigma = \phi \circ \tau \circ \phi^{-1}$ and $w = \phi(v)$ in [2].
\end{dfn}

\begin{thm}\rm
 There are exactly 14 groups of order 16 up to isomorphism. Besides the outsider $G_{0} = C_{2} \times C_{2} \times C_{2} \times C_{2}$ they can be listed as follows,
\end{thm}
\begin{center}
 \begin{tabular}{ll}
 $G_{1} = C_{2} \times C_{8}$ & ($C_{8}, 2, \sigma_{1}, e$) \\
 $G_{2} = C_{2} \ltimes_{\sigma_{2}} C_{8}$ & ($C_{8}, 2, \sigma_{2}, e$) \\
 $G_{3} = C_{2} \ltimes_{\sigma_{3}} C_{8}$ & ($C_{8}, 2, \sigma_{3}, e$) \\
 $G_{4} = C_{2} \ltimes_{\sigma_{4}} C_{8}$ & ($C_{8}, 2, \sigma_{4}, e$) \\
 $G_{5} = Q_{16}$                          & ($C_{8}, 2, \sigma_{4}, x^4$) \\
 $G_{6} = C_{16}$                          & ($C_{8}, 2, \sigma_{1}, x$) \\
 $G_{7} = K_{4} \times C_{4}$          & ($K_{8}, 2, \psi_{1}, e$) \\
 $G_{8} = D_{8} \times C_{2}$          & ($K_{8}, 2, \psi_{3}, e$) \\
 $G_{9} = C_{4} \ltimes_{\tau} K_{4}$ & ($K_{8}, 2, \psi_{5}, e$) \\              
 $G_{10} = C_{2} \ltimes_{\tau} Q_{8}$ & ($K_{8}, 2, \psi_{6}, e$) \\              
 $G_{11} = C_{2} \times Q_{8}$ & ($K_{8}, 2, \psi_{3}, x^2$) \\              
 $G_{12} = C_{4} \ltimes_{\tau} C_{4}$ & ($K_{8}, 2, \psi_{5}, x^2$) \\
 $G_{13} = C_{4} \times C_{4}$ & ($K_{8}, 2, \psi_{1}, y$) \\
 \end{tabular}
\end{center}

Refer to in [2] for the proof of Theorem 3.

\subsection{Type 1 Gray map for $G_{1}$.}
$\textbf{Example 5.1.}$ Let $G$ be $G_{1}(C_{8}, 2, \sigma_{1}, e$)         . Assume that $H$ = $C_{8} \le G$ be the maximal subgroup of $G$.  Let $\phi_{1} : H \rightarrow \mathbb{Z}^4_{2}$ be the previously constructed Gray map for $C_{8}$. Set $\phi_{2} \stackrel{\mathrm{def.}}{=} \hat{\phi}_{1}$, we have,
\begin{center}
 \begin{tabular}{lll}
 $\phi_{2}(e)$    & = ($\phi_{1}(e)\mid \phi_{1}(e)$) &  = 00000000 \\
 $\phi_{2}(x)$    & = ($\phi_{1}(x)\mid \phi_{1}(x)$) &  = 00110011 \\ 
$\phi_{2}(x^2)$ & = ($\phi_{1}(x^2)\mid \phi_{1}(x^2)$) &= 01010101 \\   
 $\phi_{2}(x^3)$ & = ($\phi_{1}(x^3)\mid \phi_{1}(x^3)$) &= 01100110 \\ 
 $\phi_{2}(x^4)$ & = ($\phi_{1}(x^4)\mid \phi_{1}(x^4)$) &= 11111111 \\ 
 $\phi_{2}(x^5)$ & = ($\phi_{1}(x^5)\mid \phi_{1}(x^5)$) &= 11001100  \\ 
$\phi_{2}(x^6)$ & = ($\phi_{1}(x^6)\mid \phi_{1}(x^6)$) &= 10101010\\ 
 $\phi_{2}(x^7)$ & = ($\phi_{1}(x^7)\mid \phi_{1}(x^7)$) &= 10011001 \\  
$\phi_{2}(a)$    &= ($\phi_{1}(e)\mid \phi_{1}(e)$ + 1111)&= 00001111 \\
$\phi_{2}(xa)$   &= ($\phi_{1}(x)\mid \phi_{1}(x)$ + 1111)&= 00111100 \\ 
$\phi_{2}(x^2a)$ &= ($\phi_{1}(x^2)\mid \phi_{1}(x^2)$ + 1111)&= 01011010 \\   
$\phi_{2}(x^3a)$ &= ($\phi_{1}(x^3)\mid \phi_{1}(x^3)$ + 1111)&= 01101001 \\ 
$\phi_{2}(x^4a)$ &= ($\phi_{1}(x^4)\mid \phi_{1}(x^4)$ + 1111) &= 11110000 \\ 
$\phi_{2}(x^5a)$ &= ($\phi_{1}(x^5)\mid \phi_{1}(x^5)$ + 1111) &= 11000011  \\ 
$\phi_{2}(x^6a)$ &= ($\phi_{1}(x^6)\mid \phi_{1}(x^6)$ + 1111) &= 10100101 \\ 
$\phi_{2}(x^7a)$ &= ($\phi_{1}(x^7)\mid \phi_{1}(x^7)$ + 1111) &= 10010110 \
 \end{tabular}
\end{center}  

\subsection{Type 1 Gray map for $G_{2},G_{3},G_{4},G_{5}$ and $G_{6}$.}
Let $G$ be $G_{2}(C_{8}, 2, \sigma_{2}, e$). Assume that $H$ = $C_{8} \le G$ be the maximal subgroup of $G$. Let $\phi_{1} : H \rightarrow \mathbb{Z}^4_{2}$ be the previously constructed Gray map for $C_{8}$. Set $\phi_{2} \stackrel{\mathrm{def.}}{=} \hat{\phi}_{1}$, we have the same Gray map with Example 5.1 for $G_{2}$. Also we have the same Gray map with Example 5.1 for $G_{3},G_{4},G_{5}$ and $G_{6}$.

\subsection{Type 1 Gray map  for $G_{7}$.}
$\textbf{Example 5.2.}$ Let $G$ be $G_{7}(K_{8}, 2, \psi_{1}, e$). Assume that $H$ = $K_{8} \le G$ be the maximal subgroup of $G$. Let $\phi_{1} : H \rightarrow \mathbb{Z}^4_{2}$ be the previously constructed Gray map for $K_{8}$. Set $\phi_{2} \stackrel{\mathrm{def.}}{=} \hat{\phi}_{1}$, we have,

\begin{center}
 \begin{tabular}{lll}
 $\phi_{2}(e)$    & = ($\phi_{1}(e)\mid \phi_{1}(e)$) & = 00000000 \\
 $\phi_{2}(x)$    & = ($\phi_{1}(x)\mid \phi_{1}(x)$) & = 00110011 \\ 
 $\phi_{2}(x^2)$ & = ($\phi_{1}(x^2)\mid \phi_{1}(x^2)$) &= 01010101 \\   
 $\phi_{2}(x^3)$ & = ($\phi_{1}(x^3)\mid \phi_{1}(x^3)$) &= 01100110 \\ 
 $\phi_{2}(y)$ & = ($\phi_{1}(y)\mid \phi_{1}(y)$) &= 11111111 \\ 
$\phi_{2}(xy)$ & = ($\phi_{1}(xy)\mid \phi_{1}(xy)$) &= 11001100  \\ 
 $\phi_{2}(x^2y)$ & = ($\phi_{1}(x^2y)\mid \phi_{1}(x^2y)$) &= 10101010\\ 
 $\phi_{2}(x^3y)$ & = ($\phi_{1}(x^3y)\mid \phi_{1}(x^3y)$) &= 10011001 \\  
$\phi_{2}(a)$    &= ($\phi_{1}(e)\mid(\phi_{1}(e)$ + 1111))&= 00001111 \\
$\phi_{2}(xa)$   &= ($\phi_{1}(x)\mid(\phi_{1}(x)$ + 1111))&= 00111100 \\ 
$\phi_{2}(x^2a)$ &= ($\phi_{1}(x^2)\mid(\phi_{1}(x^2)$ + 1111))&= 01011010 \\   
$\phi_{2}(x^3a)$ &= ($\phi_{1}(x^3)\mid(\phi_{1}(x^3)$ + 1111))&= 01101001 \\ 
$\phi_{2}(ya)$ &= ($\phi_{1}(y)\mid(\phi_{1}(y)$ + 1111)) &= 11110000 \\ 
$\phi_{2}(xya)$ &= ($\phi_{1}(xy)\mid(\phi_{1}(xy)$ + 1111)) &= 11000011  \\ 
$\phi_{2}(x^2ya)$ &= ($\phi_{1}(x^2y)\mid(\phi_{1}(x^2y)$ + 1111)) &= 10100101 \\ 
$\phi_{2}(x^3ya)$ &= ($\phi_{1}(x^3y)\mid(\phi_{1}(x^3y)$ + 1111)) &= 10010110 \
 \end{tabular}
\end{center}  

\subsection{Type 1 Gray map for $G_{8},G_{9},G_{10},G_{11}, G_{12}$ and $G_{13}$.}
Let $G$ be $G_{8}(K_{8}, 2, \psi_{3}, e$). Assume that $H$ = $K_{8} \le G$ be the maximal subgroup of $G$. Let $\phi_{1} : H \rightarrow \mathbb{Z}^4_{2}$ be the previously constructed Gray map for $K_{8}$. Set $\phi_{2} \stackrel{\mathrm{def.}}{=} \hat{\phi}_{1}$, we have the same Gray map with Example 5.2 for $G_{8}$. Also we have the same Gray map with Example 5.2 for $G_{9},G_{10},G_{11},G_{12}$ and $G_{13}$.

\section{Type 2 Gray maps for a group of order 16}
\subsection{Type 2 Gray map for $G_{1}$,$G_{2},G_{3},G_{4},G_{5}$ and $G_{6}$.}

We can not construct a Type 2 Gray map for $G_{1}$,$G_{2},G_{3},G_{4},G_{5}$ and $G_{6}$ which are semidirect products of $C_{2}$ and $C_{8}$, because we can not construct a Type 2 Gray map for $C_{8}$. (Section 4)

\subsection{Type 2 Gray map for $G_{7}$.} 
$\textbf{Example 6.1.}$ Let $G$ be $G_{7}$. We have $G\simeq C_{2} \times K_{8}$ and $v = a^2 = e$. Consider the natural Gray map $\phi_{0} : C_{2} \rightarrow \mathbb{Z}_{2}$ and the Gray map $\phi_{2} : K_{8} \rightarrow \mathbb{Z}^3_{2}$ described in the example 4.2. Now we can construct a Type 2 Gray map $\theta$ from $G_{7}$ to $\mathbb{Z}^4_{2}$ as follows,
\begin{center}
 \begin{tabular}{lll}
 $\theta(e)$    & = ($\phi_{0}(e)\mid \phi_{2}(e)$) &      = 0000 \\
 $\theta(x)$    & = ($\phi_{0}(e)\mid \phi_{2}(x)$) &      = 0001 \\ 
 $\theta(x^2)$ & = ($\phi_{0}(e)\mid \phi_{2}(x^2)$) &   = 0011 \\   
 $\theta(x^3)$ & = ($\phi_{0}(e)\mid \phi_{2}(x^3)$) &   = 0010 \\ 
 $\theta(y)$    & = ($\phi_{0}(e)\mid \phi_{2}(y)$) &     = 0100 \\ 
 $\theta(xy)$   & = ($\phi_{0}(e)\mid \phi_{2}(xy)$) &   = 0101 \\ 
 $\theta(x^2y)$ & = ($\phi_{0}(e)\mid \phi_{2}(x^2y)$) & = 0111 \\ 
 $\theta(x^3y)$ & = ($\phi_{0}(e)\mid \phi_{2}(x^3y)$) & = 0110 \\     
 $\theta(a)$     & = ($\phi_{0}(a)\mid \phi_{2}(e)$) &      = 1000 \\
 $\theta(xa)$   & = ($\phi_{0}(a)\mid \phi_{2}(x)$) &       = 1001 \\ 
 $\theta(x^2a)$ & = ($\phi_{0}(a)\mid \phi_{2}(x^2)$) &    = 1011 \\   
 $\theta(x^3a)$ & = ($\phi_{0}(a)\mid \phi_{2}(x^3)$) &    = 1010 \\ 
 $\theta(ya)$   & = ($\phi_{0}(a)\mid \phi_{2}(y)$)   &     = 1100 \\ 
 $\theta(xya)$ &  = ($\phi_{0}(a)\mid \phi_{2}(xy)$) &     = 1101 \\ 
 $\theta(x^2ya)$ & = ($\phi_{0}(a)\mid \phi_{2}(x^2y)$) & = 1111 \\ 
 $\theta(x^3ya)$ & = ($\phi_{0}(a)\mid \phi_{2}(x^3y)$) & = 1110 \\ 
 \end{tabular}
\end{center}
In $G_{7}$, we have $(xa)^{-1} = x^{3}a, (xya)^{-1} = x^3ya$. So (2) and (3) of Lemma 1 hold. 
Since in $G_{7} \simeq C_{2} \times K_{8}$,  we have $w_{H}(\phi_{2}(k)) = w_{H}(\phi_{2}(\psi_{h}(k)))$ for all h $\in C_{2}(=\{e, a\})$, and therefore $\theta$ is a Type 2 Gray map for $G_{7}$ by Theorem 2.  We can construct a Type 2 Gray map for $G_{7}$.

\subsection{Type 2 Gray map for $G_{8}$.} 
Let $G$ be $G_{8}$. We have $G\simeq C_{2} \ltimes_{\psi_{3}}  K_{8}$ and $v = a^2 = e$. Similarly, consider the natural Gray map $\phi_{0} : C_{2} \rightarrow \mathbb{Z}_{2}$ and the Gray map $\phi_{2} : K_{8} \rightarrow \mathbb{Z}^3_{2}$ described in the Example 4.2. Now we can construct a similar Type 2 Gray map as in Example 6.1.  The orders of all elements in $G_{8} \setminus K_{8}$ are 2, so (2) and (3) of Lemma 1 hold. Since in $G_{8} \simeq C_{2} \ltimes_{\psi_{3}} K_{8}$, where $ax = x^3a$ and $ay = ya$, we have $w_{H}(\phi_{2}(\psi_{a}(x))) = w_{H}(\phi_{2}(axa^{-1}) = w_{H}(\phi_{2}(x^3) = w_{H}(\phi_{2}(x))$ and  $w_{H}(\phi_{2}(\psi_{a}(xy))) = w_{H}(\phi_{2}(axya^{-1}) = w_{H}(\phi_{2}(x^3y) = w_{H}(\phi_{2}(xy))$ and so on. Hence we have $w_{H}(\phi_{2}(k)) = w_{H}(\phi_{2}(\psi_{h}(k)))$ for all $h \in C_{2}(=\{e, a\})$, and therefore $\theta$ is a Type 2 Gray map for $G_{8}$ by Theorem 2. We can construct a Type 2 Gray map for $G_{8}$.

\subsection{Type 2 Gray map for $G_{8}(D_{8}, 2, \alpha_{5}, e$)}
In [6], $D_{8}$ is one of the maximal subgroups of $G_{8}$. So, we consider to construct a Type 2 Gray map for $G_{8}$ with $N = D_{8}$ instead of $N = K_{8}$.

Let $G$ be $G_{8}(D_{8}, 2, \alpha_{5}, e$). We have $G\simeq C_{2} \ltimes_{\alpha_{5}}  D_{8}$. Consider the natural Gray map $\phi_{0} : C_{2} \rightarrow \mathbb{Z}_{2}$ and the Gray map $\phi_{2} : D_{8} \rightarrow \mathbb{Z}^3_{2}$ described in the Example 4.1. Now we can construct a similar Type 2 Gray map as in  Example 6.1.  In $G_{8}$ we have
$(xya)^{-1} = x^3ya$, so (2) and (3) of Lemma 1 hold.
 Since in $G_{8} \simeq C_{2} \ltimes_{\alpha_{5}} D_{8}$, where $ax = x^3a$ and $ay = ya$, so we have $w_{H}(\phi_{2}(\psi_{a}(x))) = w_{H}(\phi_{2}(axa^{-1}) = w_{H}(\phi_{2}(x^3) = w_{H}(\phi_{2}(x))$ and  $w_{H}(\phi_{2}(\psi_{a}(xy))) = w_{H}(\phi_{2}(axya^{-1}) = w_{H}(\phi_{2}(x^3y) = w_{H}(\phi_{2}(xy))$ and so on. Hence we have $w_{H}(\phi_{2}(k)) = w_{H}(\phi_{2}(\psi_{h}(k)))$ for all h $\in C_{2}(=\{e, a\})$, and therefore $\theta$ is a Type 2 Gray map for $G_{8}$ by Theorem 2. We can construct a Type 2 Gray map for $G_{8}$.

\subsection{Type 2 Gray map for $G_{9}$.}  
Let $G$ be $G_{9}$. We have $G\simeq C_{2} \ltimes_{\psi_{5}}  K_{8}$ and $v = a^2 = e$. Similarly, consider the natural Gray map $\phi_{0} : C_{2} \rightarrow \mathbb{Z}_{2}$ and the Gray map $\phi_{2} : K_{8} \rightarrow \mathbb{Z}^3_{2}$ described in the Example 4.2. Now we can construct a similar Type 2 Gray map as in  Example 6.1.  In $G_{9}$ we have
$(xa)^{-1} = x^{3}ya$ and $(x^3a)^{-1} = xya$, so we have $w_{H}(\theta(g)) \neq w_{H}(\theta(g^{-1}$)). Since $\theta$ does not satisfy (2) of Lemma 1, we can not construct a Type 2 Gray map for $G_{9}$.

\subsection{Type 2 Gray map for $G_{10}$.}
Let $G$ be $G_{10}$. We have $G\simeq C_{2} \ltimes_{\psi_{6}}  K_{8}$ and $v = a^2 = e$. Similarly, consider the natural Gray map $\phi_{0} : C_{2} \rightarrow \mathbb{Z}_{2}$ and the Gray map $\phi_{2} : K_{8} \rightarrow \mathbb{Z}^3_{2}$ described in the Example 4.2. Now we can construct the a similar Type 2 Gray map as in  Example 6.1.  In $G_{10}$ we have
$(xy)^{-1} = x^3y, (ya)^{-1} = x^{2}ya$ and $(xya)^{-1} = x^3ya$, so we have $w_{H}(\theta(g)) \neq w_{H}(\theta(g^{-1}$)). Since $\theta$ does not satisfy (2) of Lemma 1, we can not construct a Type 2 Gray map for $G_{10}$.

\subsection{Type 2 Gray map for $G_{10}(D_{8}, 2, \alpha_{3}, e$)}
In [6], $D_{8}$ is one of the maximal subgroups of $G_{10}$. So we consider to construct a Type 2 Gray map for $G_{10}$ with $N = D_{8}$ instead of $N = K_{8}$.

Let $G$ be $G_{10}(D_{8}, 2, \alpha_{3}, e$). We have $G\simeq C_{2} \ltimes_{\alpha_{3}}  D_{8}$. Consider the natural Gray map $\phi_{0} : C_{2} \rightarrow \mathbb{Z}_{2}$ and the Gray map $\phi_{2} : D_{8} \rightarrow \mathbb{Z}^3_{2}$ described in the Example 4.1. Now we can construct a similar Type 2 Gray map as in  Example 6.1.  In $G_{8}$ we have   
$(xa)^{-1} = x^3a, (ya)^{-1} = x^2ya$ and $(xya)^{-1} = x^3ya$, so we have $w_{H}(\theta(g)) \neq w_{H}(\theta(g^{-1}$)). Since $\theta$ does not satisfy (2) of Lemma 1, we can not construct a Type 2 Gray map for $G_{10}$.

\subsection{Type 2 Gray map for $G_{11}$.}
Let $G$ be $G_{11}$. We have $G\simeq C_{2} \ltimes_{\psi_{3}}  K_{8}$ and $v = a^2 = x^2$. Similarly, consider the natural Gray map $\phi_{0} : C_{2} \rightarrow \mathbb{Z}_{2}$ and the Gray map $\phi_{2} : K_{8} \rightarrow \mathbb{Z}^3_{2}$ described in the Example 4.2. Now we can construct a similar Type 2 Gray map as in  Example 6.1.  In $G_{11}$ we have
$a^{-1} = x^2a, (xa)^{-1} = x^{3}a, (ya)^{-1} = x^{2}ya$ and $(xya)^{-1} = x^{3}ya$, so we have $w_{H}(\theta(g)) \neq w_{H}(\theta(g^{-1}$)). Since $\theta$ does not satisfy (2) of Lemma 1,  we can not construct a Type 2 Gray map for  $G_{11}$.

\subsection{Type 2 Gray map for $G_{12}$.}
Let $G$ be $G_{12}$. We have $G\simeq C_{2} \ltimes_{\psi_{5}}  K_{8}$ and $v = a^2 = x^2$. Similarly, consider the natural Gray map $\phi_{0} : C_{2} \rightarrow \mathbb{Z}_{2}$ and the Gray map $\phi_{2} : K_{8} \rightarrow \mathbb{Z}^3_{2}$ described in the Example 4.2. Now we can construct a similar Type 2 Gray map as in  Example 6.1.   In $G_{12}$ we have
$a^{-1} = x^{2}a, (xa)^{-1} = xya, (x^3a)^{-1} = x^{3}ya$ and $(ya)^{-1} = x^2ya$, so we have $w_{H}(\theta(g)) \neq w_{H}(\theta(g^{-1}$)). Since $\theta$ does not satisfy (2) of Lemma 1,  we can not construct a Type 2 Gray map for  $G_{12}$.

\subsection{Type 2 Gray map for $G_{13}$.}
Let $G$ be $G_{13}$. We have $G\simeq C_{2} \ltimes_{\psi_{1}}  K_{8}$ and $v = a^2 = y$. Similarly, consider the natural Gray map $\phi_{0} : C_{2} \rightarrow \mathbb{Z}_{2}$ and the Gray map $\phi_{2} : K_{8} \rightarrow \mathbb{Z}^3_{2}$ described in the Example 4.2. Now we can construct a similar Type 2 Gray map as in  Example 6.1.   In $G_{13}$ we have
$a^{-1} = ya, (xa)^{-1} = x^3ya, (x^2a)^{-1} = x^2ya$ and  $(x^3a)^{-1} = xya$, so we have $w_{H}(\theta(g)) \neq w_{H}(\theta(g^{-1}$)). Since $\theta$ does not satisfy (2) of Lemma 1,  we can not construct a Type 2 Gray map for  $G_{13}$.

\section{The extension type ($C_{4},4,\tau,v$) or  ($K_{4},4,\tau,v$)}
In Theorem 3 we have $G_{7} = C_{4} \times K_{4}$, $G_{9} = C_{4} \ltimes_{\tau} K_{4}$, $G_{12} = C_{4} \ltimes_{\tau} C_{4}$ and $G_{13} = C_{4} \times C_{4}$, so we consider to construct a Type 2 Gray map for $G$ with $N = C_{4}$ or $N = K_{4}$. 

\subsection{Type 2 Gray map for $G_{7}(K_{4}, 4, \phi_{1}, e$)}
We assume that $K_{4}$ is $\langle a \rangle \times \langle y \rangle$ and $C_{4}$ is $\langle x \rangle$, where $x^4 = y^2 = a^2 = e$. 
 Let $G$ be $G_{7}$. We have $G \simeq C_{4} \times K_{4}$ and $v = x^4 = e$. Consider the Gray map $\phi_{1} : C_{4} \rightarrow \mathbb{Z}^2_{2}$ and the Gray map $\phi_{2} : K_{4} \rightarrow \mathbb{Z}^2_{2}$ described in the example 2.1 and 2.2. Now we can construct a Type 2 Gray map $\theta$ from $G_{7}$ to $\mathbb{Z}^4_{2}$ as follows,
\begin{center}
 \begin{tabular}{lll}
 $\theta(e)$    & = ($\phi_{1}(e)\mid \phi_{2}(e)$) &      = 0000 \\
 $\theta(a)$    & = ($\phi_{1}(e)\mid \phi_{2}(a)$) &      = 0001 \\ 
 $\theta(y)$ & = ($\phi_{1}(e)\mid \phi_{2}(y)$) &         = 0011 \\   
 $\theta(ya)$ & = ($\phi_{1}(e)\mid \phi_{2}(ya)$) &      = 0010 \\ 
 $\theta(x)$    & = ($\phi_{1}(x)\mid \phi_{2}(e)$) &      = 0100 \\ 
 $\theta(xa)$   & = ($\phi_{1}(x)\mid \phi_{2}(a)$) &      = 0101 \\ 
 $\theta(xy)$ & = ($\phi_{1}(x)\mid \phi_{2}(y)$) &         = 0111 \\ 
 $\theta(xya)$ & = ($\phi_{1}(x)\mid \phi_{2}(ya)$) &      = 0110 \\     
 $\theta(x^2)$     & = ($\phi_{1}(x^2)\mid \phi_{2}(e)$) & = 1100 \\
 $\theta(x^2a)$   & = ($\phi_{1}(x^2)\mid \phi_{2}(a)$) &  = 1101 \\ 
 $\theta(x^2y)$ & = ($\phi_{1}(x^2)\mid \phi_{2}(y)$) &    = 1111 \\   
 $\theta(x^2ya)$ & = ($\phi_{1}(x^2)\mid \phi_{2}(ya)$) & = 1110 \\ 
 $\theta(x^3)$   & =  ($\phi_{1}(x^3)\mid \phi_{2}(e)$)  &  = 1000 \\ 
 $\theta(x^3a)$ &  = ($\phi_{1}(x^3)\mid \phi_{2}(a)$) &  = 1001 \\ 
 $\theta(x^3y)$ & = ($\phi_{1}(x^3)\mid \phi_{2}(y)$) &    = 1011 \\ 
 $\theta(x^3ya)$ & = ($\phi_{1}(x^3)\mid \phi_{2}(ya)$) & = 1010 \\ 
 \end{tabular}
\end{center}
In $G_{7}$ we have $(xa)^{-1} = x^{3}a, (xy)^{-1} = x^3y$ and   $(xya)^{-1} = x^3ya$, so (2) and (3) of Lemma 1 hold. Since in $G_{7} \simeq C_{4} \times K_{4}$, we have $w_{H}(\phi_{2}(k)) = w_{H}(\phi_{2}(\psi_{h}(k)))$ for all h $\in C_{4}(=\{e, x, x^2, x^3\})$,
and therefore $\theta$ is a Type 2 Gray map for $G_{7}$ by Theorem 2. We can construct a Type 2 Gray map for $G_{7}$.

\subsection{Type 2 Gray map for $G_{9}(K_{4}, 4, \sigma, e$)}
In [2], $G_{9} \simeq \langle x \rangle \ltimes_{\tau}  \langle y,a \rangle = C_{4} \ltimes_{\tau} K_{4}$, where $\tau(y) = y$ and $\tau(a) = ya$ (i.e. $x(y)x^{-1} = y$ and $x(a)x^{-1} = ya$). Because all order two elements $\sigma$ of Aut($K_{4}) \simeq S_{3}$ are conjugate, they yield isomorphic semidirect products $C_{4} \ltimes_{\sigma} K_{4}$.

 Let $G$ be $G_{9}$. We have $G \simeq C_{4} \ltimes_{\sigma} K_{4}$ and $v = x^4 = e$. Consider the Gray map $\phi_{1} : C_{4} \rightarrow \mathbb{Z}^2_{2}$ and the Gray map $\phi_{2} : K_{4} \rightarrow \mathbb{Z}^2_{2}$ described in the Example 2.1 and 2.2. Now we can construct a Type 2 Gray map $\theta$ from $G_{9}$ to $\mathbb{Z}^4_{2}$ as follows,
\begin{center}
 \begin{tabular}{lll}
 $\theta(e)$    & = ($\phi_{1}(e)\mid \phi_{2}(e)$) &      = 0000 \\
 $\theta(a)$    & = ($\phi_{1}(e)\mid \phi_{2}(a)$) &      = 0001 \\ 
 $\theta(y)$ & = ($\phi_{1}(e)\mid \phi_{2}(y)$) &         = 0011 \\   
 $\theta(ya)$ & = ($\phi_{1}(e)\mid \phi_{2}(ya)$) &      = 0010 \\ 
 $\theta(x)$    & = ($\phi_{1}(x)\mid \phi_{2}(e)$) &      = 0100 \\ 
 $\theta(xa)$   & = ($\phi_{1}(x)\mid \phi_{2}(a)$) &      = 0101 \\ 
 $\theta(xy)$ & = ($\phi_{1}(x)\mid \phi_{2}(y)$) &         = 0111 \\ 
 $\theta(xya)$ & = ($\phi_{1}(x)\mid \phi_{2}(ya)$) &      = 0110 \\     
 $\theta(x^2)$     & = ($\phi_{1}(x^2)\mid \phi_{2}(e)$) & = 1100 \\
 $\theta(x^2a)$   & = ($\phi_{1}(x^2)\mid \phi_{2}(a)$) &  = 1101 \\ 
 $\theta(x^2y)$ & = ($\phi_{1}(x^2)\mid \phi_{2}(y)$) &    = 1111 \\   
 $\theta(x^2ya)$ & = ($\phi_{1}(x^2)\mid \phi_{2}(ya)$) & = 1110 \\ 
 $\theta(x^3)$   & = ($\phi_{1}(x^3)\mid \phi_{2}(e)$)  &  = 1000 \\ 
 $\theta(x^3a)$ &  = ($\phi_{1}(x^3)\mid \phi_{2}(a)$) &  = 1001 \\ 
 $\theta(x^3y)$ & = ($\phi_{1}(x^3)\mid \phi_{2}(y)$) &    = 1011 \\ 
 $\theta(x^3ya)$ & = ($\phi_{1}(x^3)\mid \phi_{2}(ya)$) & = 1010 \\ 
 \end{tabular}
\end{center}
In $G_{9}$ we have $(xy)^{-1} = x^{3}y, (xa)^{-1} = x^3ya$ and   $(x^3a)^{-1} = xya$, so (2) and (3) of Lemma 1 hold. Since $G_{9} \simeq C_{4} \ltimes_{\sigma} K_{4}$, where $xy = yx$ and $xa = yax$, so we have $w_{H}(\phi_{2}(\psi_{x}(a))) = w_{H}(\phi_{2}(xax^{-1}) = w_{H}(\phi_{2}(ya) = w_{H}(\phi_{2}(a))$ and  $w_{H}(\phi_{2}(\psi_{x^2}(a))) = w_{H}(\phi_{2}(x^2ax^2) = w_{H}(\phi_{2}(a))$ and so on. Hence we have $w_{H}(\phi_{2}(k)) = w_{H}(\phi_{2}(\psi_{h}(k)))$ for all h $\in C_{4}(=\{e, x, x^2, x^3\})$, and therefore $\theta$ is a Type 2 Gray map for $G_{9}$ by Theorem 2. We can construct a Type 2 Gray map for $G_{9}$.

\subsection{Type 2 Gray map for $G_{12}(C_{4}, 4, \varphi_{2}, e$)}
Similarly $G_{12} \simeq \langle x \rangle \ltimes_{\tau}  \langle xa \rangle = C_{4} \ltimes_{\tau} C_{4}$, where $\langle xa \rangle = \{e, xa, y, xya\}, \tau = \varphi_{2}$ (i.e. $\varphi_{2}(xa) = xya$) and $v = a^4 = e$. Consider the Gray map $\phi_{1} : C_{4}(=\langle x \rangle) \rightarrow \mathbb{Z}^2_{2}$ and the Gray map $\phi_{2} : C_{4}(=\langle xa \rangle) \rightarrow \mathbb{Z}^2_{2}$ described in the Example 2.1. Now we can construct a Type 2 Gray map $\theta$ from $G_{12}$ to $\mathbb{Z}^4_{2}$ as follows,
\begin{center}
 \begin{tabular}{lll}
 $\theta(e)$    &  = ($\phi_{1}(e)\mid \phi_{2}(e)$) &      = 0000 \\
 $\theta(xa)$    & = ($\phi_{1}(e)\mid \phi_{2}(xa)$) &     = 0001 \\ 
 $\theta(y)$ &     = ($\phi_{1}(e)\mid \phi_{2}(y)$) &       = 0011 \\   
 $\theta(xya)$ &  = ($\phi_{1}(e)\mid \phi_{2}(xya)$) &   = 0010 \\ 
 $\theta(x)$    &  = ($\phi_{1}(x)\mid \phi_{2}(e)$) &      = 0100 \\ 
 $\theta(x^2a)$ & = ($\phi_{1}(x)\mid \phi_{2}(xa)$) &      = 0101 \\ 
 $\theta(xy)$ &     = ($\phi_{1}(x)\mid \phi_{2}(y)$) &    = 0111 \\ 
 $\theta(x^2ya)$ & = ($\phi_{1}(x)\mid \phi_{2}(xya)$) & = 0110 \\     
 $\theta(x^2)$ &    = ($\phi_{1}(x^2)\mid \phi_{2}(e)$) &  = 1100 \\
 $\theta(x^3a)$ &   = ($\phi_{1}(x^2)\mid \phi_{2}(xa)$) & = 1101 \\ 
 $\theta(x^2y)$ &   = ($\phi_{1}(x^2)\mid \phi_{2}(y)$) & = 1111 \\   
$\theta(x^3ya)$ & = ($\phi_{1}(x^2)\mid \phi_{2}(xya)$) & = 1110 \\ 
 $\theta(x^3)$   & = ($\phi_{1}(x^3)\mid \phi_{2}(e)$)  &  = 1000 \\ 
 $\theta(a)$ &      = ($\phi_{1}(x^3)\mid \phi_{2}(xa)$) &  = 1001 \\ 
 $\theta(x^3y)$ &   = ($\phi_{1}(x^3)\mid \phi_{2}(y)$) & = 1011 \\ 
 $\theta(ya)$ &   = ($\phi_{1}(x^3)\mid \phi_{2}(xya)$) & = 1010 \\ 
 \end{tabular}
\end{center}
In $G_{12}$ we have $(xy)^{-1} = x^3y, a^{-1} = x^2a, (xa)^{-1} = xya, x^3a^{-1} = x^3ya$ and $(ya)^{-1}= x^2ya$, so (2) and (3) of Lemma 1 hold.
Since in $G_{12} \simeq C_{4} \ltimes_{\varphi_{2}} C_{4}$, where $\varphi_{2}(xa) = xya$, so we have $w_{H}(\phi_{2}(\psi_{x}(xa))) = w_{H}(\phi_{2}(xya) = w_{H}(\phi_{2}(xa)$ and so on. Hence  we have $w_{H}(\phi_{2}(k)) = w_{H}(\phi_{2}(\psi_{h}(k)))$ for all $h \in C_{4}(=\{e, x, x^2, x^3\})$, and therefore $\theta$ is a Type 2 Gray map for $G_{12}$ by Theorem 2. We can construct a Type 2 Gray map for $G_{12}$.

\subsection{Type 2 Gray map for $G_{13}(C_{4}, 4, \varphi_{1}, e$)}
In [2], $G_{13} \simeq C_{4}(=\langle x \rangle) \times C_{4}(=  \langle a \rangle)$, where $\langle a \rangle = \{e, a, y, ya\}, \tau(a) = \varphi_{1}(a) = a$ (namely $xa = ax)$ and $v = a^4 = y^2 = e$.   Consider the Gray map $\phi_{1} : C_{4}(=\langle x \rangle) \rightarrow \mathbb{Z}^2_{2}$ and the Gray map $\phi_{2} : C_{4}(=\langle a \rangle) \rightarrow \mathbb{Z}^2_{2}$ described in the example 2.1. Now we can construct a Type 2 Gray map $\theta$ from $G_{13}$ to $\mathbb{Z}^4_{2}$ as follows,
\begin{center}
 \begin{tabular}{lll}
 $\theta(e)$    & = ($\phi_{1}(e)\mid \phi_{2}(e)$) &   = 0000 \\
 $\theta(a)$    & = ($\phi_{1}(e)\mid \phi_{2}(a)$) &   = 0001 \\ 
 $\theta(y)$    & = ($\phi_{1}(e)\mid \phi_{2}(y)$) &   = 0011 \\   
 $\theta(ya)$   & = ($\phi_{1}(e)\mid \phi_{2}(ya)$) &  = 0010 \\ 
 $\theta(x)$  &   = ($\phi_{1}(x)\mid \phi_{2}(e)$) &   = 0100 \\ 
 $\theta(xa)$  & = ($\phi_{1}(x)\mid \phi_{2}(a)$) &    = 0101 \\ 
 $\theta(xy)$  &  = ($\phi_{1}(x)\mid \phi_{2}(y)$) &    = 0111 \\ 
 $\theta(xya)$  & = ($\phi_{1}(x)\mid \phi_{2}(ya)$) &  = 0110 \\     
 $\theta(x^2)$    & = ($\phi_{1}(x^2)\mid \phi_{2}(e)$)  & = 1100 \\
 $\theta(x^2a)$  & = ($\phi_{1}(x^2)\mid \phi_{2}(a)$) &   = 1101 \\ 
 $\theta(x^2y)$  & = ($\phi_{1}(x^2)\mid \phi_{2}(y)$)   & = 1111 \\   
 $\theta(x^2ya)$ & = ($\phi_{1}(x^2)\mid \phi_{2}(ya)$) & = 1110 \\ 
 $\theta(x^3)$  &  = ($\phi_{1}(x^3)\mid \phi_{2}(e)$)  & = 1000 \\ 
 $\theta(x^3a)$ & = ($\phi_{1}(x^3)\mid \phi_{2}(a)$) &   = 1001 \\ 
 $\theta(x^3y)$ &  = ($\phi_{1}(x^3)\mid \phi_{2}(y)$) &   = 1011 \\ 
 $\theta(x^3ya)$ & = ($\phi_{1}(x^3)\mid \phi_{2}(ya)$) & = 1010 \\ 
 \end{tabular}
\end{center}
In $G_{13}$, we have $a^{-1} = ya, x^{-1} = x^3, (xa)^{-1} = x^3ya,  (xy)^{-1} = x^3y, (xya)^{-1} = x^3a$ and $(x^2a)^{-1} = x^2ya$, so (2) and (3) of Lemma 1 hold. Since in $G_{13} \simeq C_{4} \times C_{4}$, we have $w_{H}(\phi_{2}(k)) = w_{H}(\phi_{2}(\psi_{h}(k)))$ for all h $\in C_{4}(=\{e, x, x^2, x^3\})$, and therefore $\theta$ is a Type 2 Gray map for $G_{13}$ by Theorem 2.
we can construct a Type 2 Gray map for $G_{13}$.

\section{Summary}
\subsection{Type 1 Gray map}
We can construct a Type 1 Gray map for $G_{0}, G_{1},\dots , G_{12}$ and $G_{13}$.
\subsection{Type 2 Gray map}
1. We can construct a Type 2 Gray map for $G_{0}, G_{7}, G_{8}, G_{9}, G_{12}$ and $G_{13}$. \\
2. A necessary conditions that we can construct a Type 2 Gray map for a group of order 16 are, \\
(1)  In the extension type ($N, 2, \tau, v$) $N$ is $K_{8}$ or $D_{8}$ or $C_{2} \times C_{2} \times C_{2}$ and $v$ is $e$. \\
(2)  Alternatively in the extension type ($N, 4, \tau, v$) $N$ is $C_{4}$ or $K_{4}$ and $v$ is $e$. \\
(3)  $G$ do not contain the subgroup $C_{8}$ or $Q_{8}$.

\begin{description}
\item{\textbf{References}}
\end{description}
\begin{enumerate}
\item Reza Sobhani, Gray isometries for finite p-groups, Transactions on Combinatorics, Vol.2 No.1(2013), pp17-26.
\item Marcel Wild, The groups of order sixteen made easy, The Mathematical Association of America, January 2005.
\item A.~R.~Hammons, P.~V.~Kummar, A.~R.~Calderbank, N.~J.~A.~Sloane and P.~Sole, The $\mathbb{Z}_{4}$ linearity of Kerdock, Preparata, Goethals and related codes, IEEE Trans. Inform. Theory, 40(1994).
\item Joseph J.~Rotman, An introduction to the Theory of Groups, Springer 1995.
\item J.~H.~van Lint, Introduction to Coding Theory, Third Edition, Springer 1999.
\item A.~D.~Thomas and G.~V.~Wood, Group Tables, Mathematics Series 2, Shiva Publishing Limited, Kent, UK, 1980.  
\item F.~J.~Macwilliams and N.~J.~A.~Sloane, The theory of error-correcting codes, North-Holland, Amsterdam, 1977. 
\end{enumerate}
\end{document}